# Magnetic sensitivity distribution of Hall devices in antiferromagnetic switching experiments

F. Schreiber[1], H. Meer[1], C. Schmitt[1], R. Ramos[2,3], E. Saitoh[2,4,5,6,7], L. Baldrati[1] and M. Kläui[1,8,a]

[1]*Institute of Physics, Johannes Gutenberg-University Mainz, 55099 Mainz, Germany*

[2]*WPI-Advanced Institute for Materials Research, Tohoku University, Sendai 980-8577, Japan*

[3]*Centro de Investigación en Química Biolóxica e Materiais Moleculares (CIQUS), Departamento de Química-Física, Universidade de Santiago de Compostela, Santiago de Compostela 15782, Spain*

[4]*Institute for Materials Research, Tohoku University, Sendai 980-8577, Japan*

[5]*Advanced Science Research Center, Japan Atomic Energy Agency, Tokai 319-1195, Japan*

[6]*Center for Spintronics Research Network, Tohoku University, Sendai 980-8577, Japan*

[7]*Department of Applied Physics, The University of Tokyo, Tokyo 113-8656, Japan*

[8]*Center for Quantum Spintronics, Department of Physics, Norwegian University of Science and Technology, Trondheim, Norway*

[a] Klaeui@Uni-Mainz.de

## Abstract

We analyze the complex impact of the local magnetic spin texture on the transverse Hall-type voltage in device structures utilized to measure magnetoresistance effects. We find a highly localized and asymmetric magnetic sensitivity in the eight-terminal geometries that are frequently used in current-induced switching experiments, for instance to probe antiferromagnetic materials. Using current-induced switching of antiferromagnetic NiO/Pt as an example, we estimate the change in the spin Hall magnetoresistance signal associated with switching events based on the domain switching patterns observed via direct imaging. This estimate correlates with the actual electrical data after subtraction of a non-magnetic contribution. Here, the consistency of the correlation across three measurement geometries with fundamentally different switching patterns strongly indicates a magnetic origin of the measured and analyzed electrical signals.

# Manuscript

## Introduction

Within the rapidly developing fields of spintronics, antiferromagnetic materials (AFMs) have attracted significant attention in recent years, due to their characteristics such as potential THz switching speeds and resilience against external fields, which are promising features for the application in data storage devices [1]. Thus, current research focuses on investigating the possibility of all-electrical control of the antiferromagnetic order in both metallic and insulating AFMs by utilizing various device structures to apply current pulses and detect changes in the resistance. Two device geometries are repeatedly used in such experiments: firstly, the four-terminal *Hall cross* structure depicted in Fig. 1 (a) [2–6], and secondly, to overcome limitations of the conventional geometry, an eight-terminal *Hall star* visible in Fig. 1 (b) [2–10]. Based on elongated cross-like structures with two transverse readout channels, the field-controlled spin reorientation in insulating AFM (iAFM)/heavy metal (HM) bilayers was detected via their spin-Hall magnetoresistance (SMR) response [11–13]. While these findings indicate the possibility of an electrical readout mechanism, combining the reliable detection of changes in the SMR signal with current-induced switching is challenging. The direct observation of current-induced domain switching by imaging demonstrates the electrical writing of the Néel order in iAFM/HM bilayers [9,10,14–17]. However, the pulses with current densities of up to $10^{12}$ A/m² applied to induce magnetic changes were found to cause structural changes in the conductive layer, contributing non-magnetic signals to the transverse resistance readout [5,6,8,18,19]. Thus, changes in the transverse resistance upon the application of current pulses are not necessarily purely the magnetoresistive signature of current-induced magnetic domain switching.

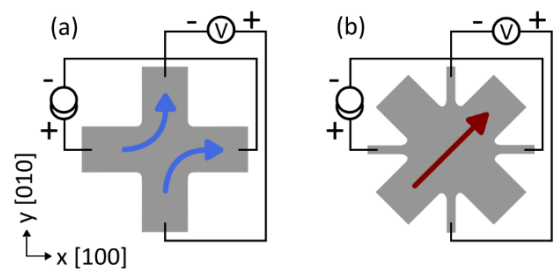

**Fig. 1**: Device layout and measurement scheme of devices utilized in antiferromagnetic switching experiments. In the four-terminal *Hall cross* in (a), the current pulses (arrows) and the readout measurement use the same channels. In the eight-terminal *Hall star* in Structural changes such as annealing or electromigration are typically promoted by elevated temperatures, which occur via increased Joule heating especially at constrictions and at the device corners [6,18,19]. The separation of pulsing and readout channels hence can possibly reduce the impact of non-magnetic effects on the transverse resistance readout. This is realized in eight-terminal *Hall star* devices, as depicted in Fig. 1 (b), which are frequently used to study both metallic AFMs [2,3] and insulating AFM/HM bilayers [4,5,7–10]. In particular for the example of NiO, the *Hall star* geometry moreover allows one to apply straight pulses along the <110> in-plane projection of the easy axes, while measuring the transverse resistance at 45° with respect to the possible in-plane components of the Néel vector $\boldsymbol{n}$, which maximizes the transverse SMR response ($\rho_\text{T} \sim n_\text{x} n_\text{y}$) [11] expected from magnetic switching.

Recently, a comparison of the two configurations displayed in Fig. 1 revealed fundamental differences in the domain switching patterns of NiO/Pt, despite an effectively identical direction of the pulses at the device center [10]. While the current at the center might flow in the same direction, there are distinctly different current density distributions for the two pulsing geometries [10], showing that care needs to be taken when comparing results obtained in different measurement configurations. Notably, the readout current experiences different geometrical confinement in the two schemes displayed in Fig. 1 as well. To quantitatively compare the typical transverse resistance measurements between these reading geometries, knowledge of the distinct magnetic sensitivity is key.

In this work, we clarify to what extent the different current distributions lead to variations in the sensitivity to local magnetic changes. We investigate the influence of the device geometry on the sensitivity of the transverse resistance readout to localized magnetic signals across the device. With an easily implemented method based on numerical simulations, we find large differences between *Hall cross* and *Hall star* device signals and apply these findings to an experimental study of current-induced switching in NiO/Pt bilayers. In the different measurement configurations, we determine the expected change in the SMR signal based on images of the domain switching and correlate this estimate with the concurrently measured electrical signals.

**Modelling sensitivity distribution for Hall signals**

To assess the sensitivity of the transverse resistance readout to local magnetic signals in a particular device structure, we simulate the current density distribution of the reading current and the transverse resistance arising from local variations in the conductivity tensor $\boldsymbol{\sigma}$ by using COMSOL Multiphysics [20]. A transverse Hall signal generally originates from the off-diagonal elements of $\boldsymbol{\sigma}$, typically in the form of a non-zero $\sigma_{xy}$. The characteristics of distinct effects such as the anisotropic magnetoresistance (AMR) or the SMR lead to quantitative differences in $\sigma_{xy}$ but obey a symmetry similar to the ordinary Hall effect (OHE). The impact of local changes in $\sigma_{xy}$ on the transverse voltage can thus be modelled in a general way, based on the OHE, to determine the restrictive effects of the device geometry on the distribution of the reading current. In our simulations, we specify the relation of $\boldsymbol{\sigma}$ to the magnetic field $B$ according to the Hall effect and determine the transverse voltage in the device while scanning the structure with a $B$ field in form of a two-dimensional Gaussian function. Mapping the transverse voltages to the corresponding regions reflects the sensitivity of the readout to magnetic signals in these areas.

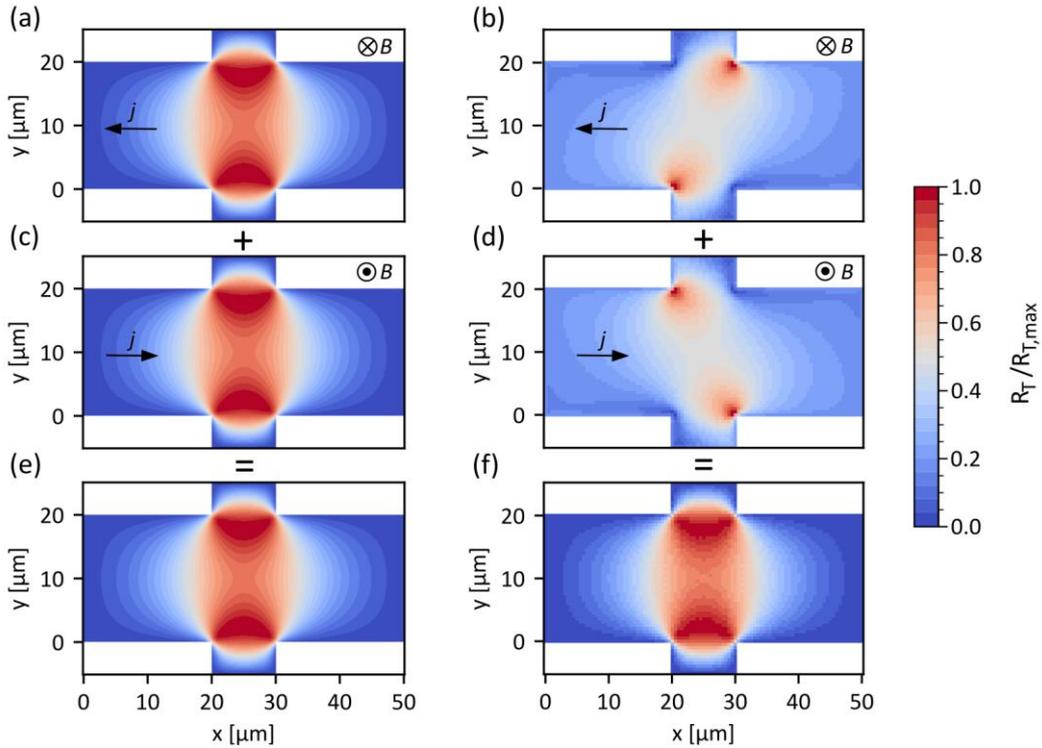

**Fig. 2**: Simulations of the local magnitude of the Hall response in a bar geometry obtained from a scan with a localized magnetic field. At Hall coefficients $R_H = -0.24 \times 10^{-10}$ m³/C (left column) and $-1 \times 10^{-5}$ m³/C (right column), the magnitude of the Hall resistance

In detail, the simulations consider a freestanding single layer of material in the shape of the corresponding Pt device. Neglecting any additional layers or the influence of magnetic order, the boundaries are specified insulating while the device material itself is set to the properties of Pt at 293.15 K from the COMSOL materials library, in particular a conductivity $\sigma_0$ = 8.9x10$^6$ S/m. The electrical connections are specified as a current source of $I_0 = 210$ μA and the corresponding ground, assigned to the two opposite edges of one channel of the device. The edges of the orthogonal channel are floating potentials $V_{1,2}$, which together yield the transverse Hall voltage $V_\text{T} = V_2 - V_1$. The *Electric Currents* interface of COMSOL determines the current distribution $j$ by solving the equation of continuity $\nabla j = -\nabla d(\boldsymbol{\sigma}\nabla V) = 0$ based on Ohm's law $j = \boldsymbol{\sigma} E$. Here, $E = -\nabla V$ is the electric field determined by the electric potential $V$ and $d$ = 2 nm is the thickness of the layer. The simulation is considered a 2D problem, i.e. the electric potential varies only in the $x$ and $y$ directions. The three-dimensional conductivity tensor $\boldsymbol{\sigma}$ is specified as

$$\boldsymbol{\sigma} = \begin{bmatrix} \sigma_{xx} & \sigma_{xy} & 0 \\ \sigma_{yx} & \sigma_{yy} & 0 \\ 0 & 0 & \sigma_0 \end{bmatrix}.$$

The tensor elements are

$$\sigma_{xx} = \sigma_{yy} = \frac{\sigma_0}{1 + (\sigma_0 R_\text{H} B)^2}$$

and

$$-\sigma_{xy} = \sigma_{yx} = \frac{\sigma_0^2 R_\text{H} B}{1 + (\sigma_0 R_\text{H} B)^2}$$

with $R_\text{H} = -0.24 \times 10^{-10}$ m³/C being the Hall coefficient of Pt [21]. Fig. 2 (a) shows the local response from scanning the whole device with a magnetic field $B$, modeled by a two-dimensional Gaussian with a full width at half maximum (FWHM) of about 1 μm and a peak level of 0.1 T. At each position $(x, y)$, the plot shows the magnitude of the Hall resistance $R_T(x, y) = V_T(x, y)/I_0$ resulting from the field $B(x, y)$. A second scan using inverse current polarity and field directions (see Fig. 2 (c)) yields a similar result such that the summed response of both simulations gives the magnetic sensitivity distribution displayed in Fig. 2 (e), where $R_T(x, y)$ is normalized to a uniform maximum level $R_{T,\max}$. Note that an increased Hall coefficient of $R_H \approx -1 \times 10^{-5}$ m³/C results in an asymmetry of the sensitivity distributions for positive and negative currents and fields, respectively, as visible in Fig. 2 (b) and (d). This likely is a result of the repulsive force from a highly increased electron density at the device edges, being significant only in case of such large $R_H$. However, by summing both plots in Fig. 2 (b) and (d), one obtains the symmetric sensitivity distribution visible in Fig. 2 (f), which is similar to the result obtained for $R_H = -0.24 \times 10^{-10}$.

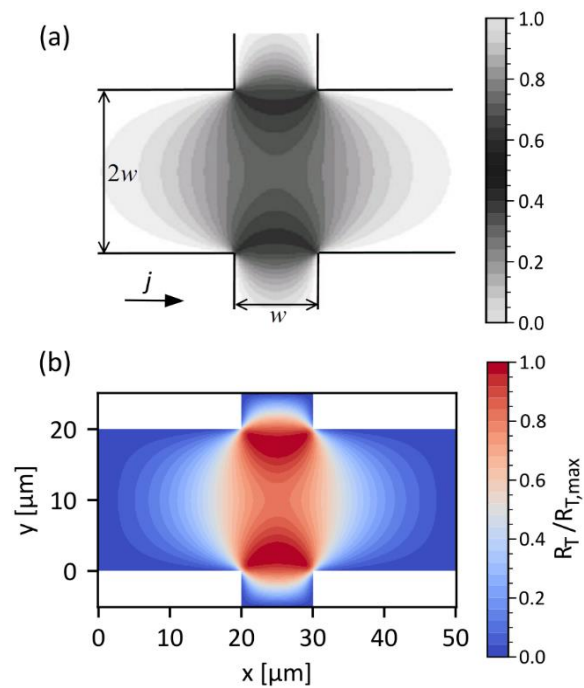

**Fig. 3**: Verification of the simulated Hall sensitivity distribution. Plot (a) shows the analytical solution for a Hall bar geometry, based on a previous work [22]. The simulations performed with COMSOL Multiphysics [20] in a similar geometry yield the magnetic sensitivity map displayed in (b), which compares to the analytical solution in (a). Plot (a) adapted from reference [23] with permission of the author.

These total sensitivity distributions of the *Hall Bar* geometry compare to the analytical solution in an equivalent geometry [22,23], as visible by comparing Fig. 3 (a) and (b). We thus conclude that summing the simulated response of two inverse field and current settings allows one to assess the sensitivity distribution of the transverse Hall voltage to local magnetic signals in the device geometry of interest.

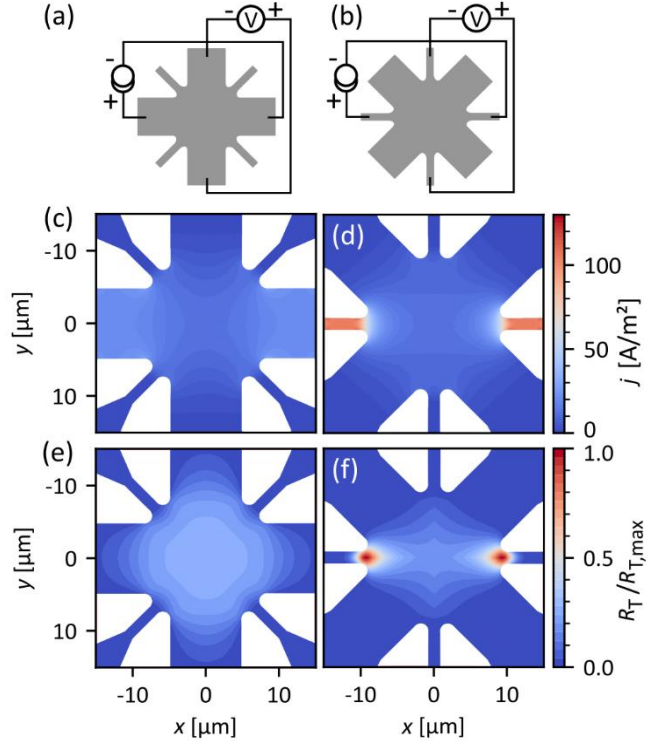

**Fig. 4**: Simulations of the readout properties of the two measurement configurations in eight-terminal Hall devices (*Hall stars*) shown in (a) and (b). The respective current density distributions of the *I* = 210 µA reading current are visible in (c) and (d). The sensitivity of the transverse resistance readout to local magnetic signals is shown in (e) and (f). Simulations were carried out using COMSOL Multiphysics [20].

While arbitrary device geometries do not allow for analytical solutions, the numerical simulations are easily transferable. We restrict the discussion to the eight-terminal *Hall stars* visible in Fig. 4, where the source current and measured voltage are connected *cross-like* (compare to Fig. 1 (a)) along the wider channels (10 µm) in Fig. 4 (a), as well as connected along the smaller channels (2 µm) in Fig. 4 (b).

Simulations of the respective current density distributions of the reading current *I* = 210 µA, without any magnetic field applied, are shown in Fig. 4 (c) and (d). In the *cross-like* measurement geometry in (c), the channel width and size of the device center are similar, therefore the current density is relatively uniform across the device center and decreases towards the voltage readout contacts. In contrast, when the readout current flows along the small channels, the current density in Fig. 4 (d) is greatly enhanced and, as it enters the cross area, sharply drops due to the comparably wide central area. This information is key to understand the sensitivity of the transverse voltage readout to magnetic signals in different regions of the device.

The magnetic sensitivity distributions of eight-terminal *Hall stars* in the two different readout configurations are shown for comparison in Fig. 4 (e) and (f), which reflect the differences in the respective current density distributions in Fig. 4 (c) and (d). The sensitivity distribution of the *cross-like Hall star* in Fig. 4 (e) is highly symmetric, features a maximum in the device center and expands to off-center regions in the individual arms. This pattern is similar to the analytical solution of a symmetric four-terminal *Hall cross* [22], which further indicates that results obtained in the *cross-like* connected eight-terminal device can be compared to measurements in four-terminal *Hall crosses* [10]. In contrast, Fig. 4 (f) reveals a strongly enhanced response of the transverse voltage to magnetic signals in the area where the small channels carrying the reading current enter the central region. The sensitivity reduces towards the device center and nearly diminishes towards the voltage reading contacts.

This non-intuitive sensitivity distribution is related to the large difference in the size of the current injection contacts with respect to the central region, leading to a larger measurement current density near the injection points. The presence and magnitude of magnetic contributions to the transverse resistance readout thus depends strongly on the position of the magnetic signal. This is crucial information if magnetic signals are localized only in certain regions or are very inhomogeneous, for example, when investigating domain switching patterns and the associated magnetoresistance changes.

Having proposed a method to understand the electrical switching signal based on the local magnetic sensitivity, we next employ this method and focus on iAFM/HM bilayers, NiO/Pt in particular, where the aforementioned device geometries were frequently used to enable the SMR readout of current-induced switching of antiferromagnetic domains [4,5,9,10,14,15,18]. NiO is a well-known antiferromagnet with collinear spin order below the bulk Néel temperature of about 523 K [24]. Furthermore, NiO exhibits a comparably large magneto-elastic coupling [25–27] and magneto-optical response [24,28], which enables imaging of the antiferromagnetic domains in NiO via magneto-optical birefringence effects [9,28]. NiO/Pt is therefore well suited to study the current-driven spin reorientation by imaging and electrical measurements on devices patterned in the Pt layer. Previous reports on the NiO domain structure often reported small domains, <1 µm in diameter, and inhomogeneous switching patterns [14–16]. The absence of magnetic signals from the NiO layer in some reports [5,18,19] may partly be related to such small domain sizes. In cases where the *Hall star* geometry (Fig. 4 (b)) with a readout along the small arms was used [5], the highly localized readout sensitivity visible in Fig. 4 (f) may additionally hamper the detection of magnetic signals, which could explain some of the surprising transport results previously published. Recently, we observed large domains in NiO/Pt thin films (up to 100 µm²) by birefringence imaging [9] and reported the correlation of a part of the electrical signal with the size of the switched domains, indicating a magnetic SMR contribution in the transverse resistance. In this work, we expand the experimental studies of MgO(001)//NiO(10nm)/Pt(2nm) to different device orientations with respect to the crystallographic directions of NiO and thereby utilize both measurement layouts discussed in Fig. 4 to maximize the expected SMR response.

**Experimental results**

The electrical measurements were carried out with three Keithley devices and a Keysight 34970A Matrix switch to automate the contact routing. A Keithley 6221 AC Current source supplies pulsing currents of tens of mA (~1x10$^{12}$ A/m²). The pulses referred to in this work are single cycle current bursts with a duration on the order of 1 ms. After a pulse, a delay of 10s is set to mitigate the influence of short-term resistivity fluctuations related to thermal effects such as Joule heating. Subsequently, a Keithley 2400 Sourcemeter supplies the reading current $I_R$ of 210 µA and measures the longitudinal device resistance $R_L$, while a Keithley 2182A Nanovoltmeter detects the transverse voltage $V_T$. To determine the transverse resistance $R_T$, averaging over two measurements with inverse current polarity is ought to minimize the influence of thermal effects like the Seebeck effect:

$$R_T = \frac{1}{2}\left[\frac{V_T(I_R^+)}{I_R^+} + \frac{V_T(I_R^-)}{I_R^-}\right].$$

Five of such measurement pairs were detected and averaged before the next pulse was applied. While performing the electrical measurements, we simultaneously investigated the domain switching patterns by magneto-optical birefringence imaging with a commercial Kerr microscope in polar mode [9,28,29].

While 12 possible domain configurations exist in bulk NiO [24], NiO/Pt thin films exploit only 4 domains with increased out-of-plane (OOP) orientation as a result of straining from the substrate [30,28,27]. In our samples, some of us recently determined the easy axes to be along the < ±5 ±5 19 > directions [27]. Of these 4 domains, imaging based on the magneto-optical birefringence effect allows to distinguish those with orthogonal in-plane (IP) components of the Néel vector $\boldsymbol{n}$ [28], which results in a two-level contrast. In the example of Fig. 5 (a), domains which look bright in the difference image are thus not necessarily equal but may feature antiparallel $\boldsymbol{n}$. However, such domains are similar in the associated SMR response as well since the transverse resistance contribution of the SMR $\rho_T$ is proportional to the product of the IP Néel vector components ($\rho_T \sim n_x n_y$) [11]. Accordingly, one can translate switched domains which appear bright in the difference image to an increase in the average SMR signal and such which appear dark, to a decrease.

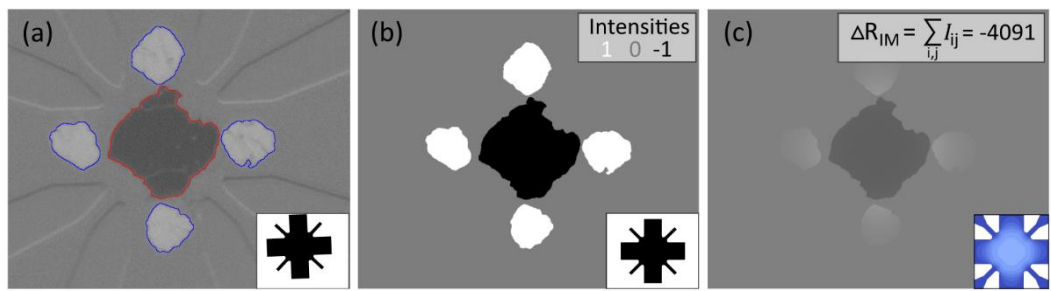

**Fig. 5**: Estimate of the SMR signal based on difference images of domain switching events. Detection of the changes visible in a difference image (a) yields the number of pixels affected by the switching. The definition of an appropriate threshold singles out the switched domains (b). After aligning the images to a uniform orientation (see insets (a) and (b)), the threshold image (b) is weighted with the Hall sensitivity distribution (inset (c)). The sum of the intensity levels visible in (c) is an estimation of the impact of the domain switching on the average SMR signal.

To estimate the relative change in the SMR signal based on difference images of the domain switching patterns, we extract the domain shapes from the difference image (Fig. 5 (a)) and create a three-level contrast image as displayed in Fig. 5 (b). In the second step, weighting by the sensitivity distribution of the corresponding measurement geometry accounts for the local variation of the impact of domain switching on the resistance readout. A thus processed image is shown in Fig. 5 (c). The sum of the weighted intensity values is an estimate of the change in the SMR signal associated to the switching event and termed $\Delta R_{\text{IM}}$.

Note that the absolute value of $\Delta R_{\text{IM}}$ includes an arbitrary scaling factor in comparison to the measured $\Delta R_{\text{T}}$. Consequently, one may always fit the response predicted from imaging to the actual transverse resistance measurement when considering a single pulse. To judge similarities and the correlation of both quantities, it is inevitable to trace the changes over several pulses.

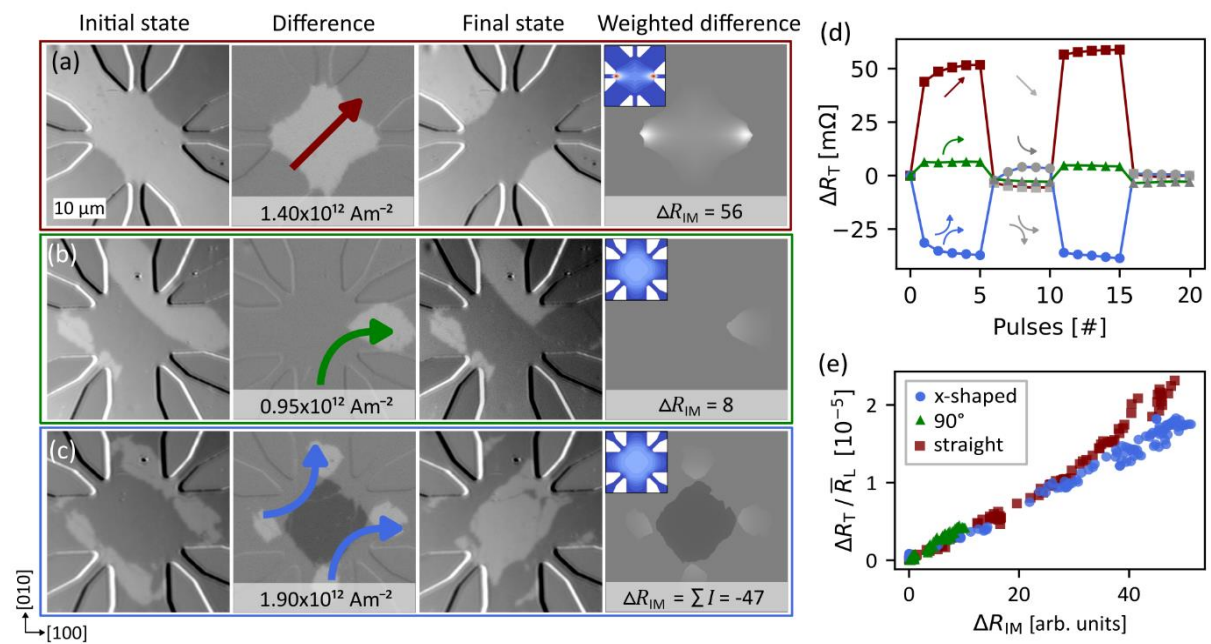

**Fig. 6**: Imaging and electrical measurements of current-induced switching in different geometries on NiO(001)/Pt. Panels (a), (b) and (c) show the changes of the domain structure caused by the application of a current pulse, as denoted by the arrows in the difference image between the initial and final state domain image. The column on the right of (a) – (c) includes the difference images after setting a threshold and weighting the difference with the magnetic sensitivity distribution of the electrical readout, displayed in the corresponding insets. The transverse resistance data in (d) was measured simultaneously and is displayed after the subtraction of a linear contribution in each pulse direction. The correlation of the estimated SMR signal change from the difference images with the actual electrical data is displayed in (e), which includes the data of multiple sequences at different pulse current densities in each geometry.

Combining the information obtained from imaging and electrical measurements, we find that current pulses on the order of 20 mA ($\sim 1 \times 10^{12}$ A/m²) and a duration of 100 µs (x-shaped and straight pulses in Fig. 6 (a) and (c), respectively) or 1 ms (90° pulses in Fig. 6 (b)) result in visible domain switching in our devices. Figs. 6 (a) – (c) include images of the domain state before and after the application of a current pulse, as indicated by the arrows in the respective difference images. One notes that the switching patterns in (a) and (c) differ greatly despite an effectively equal direction of the current pulse along [110] in the center of the device. This has been reported to be a result of differences in the thermal expansion and the related strain, which induces the spin reorientation via magneto-elastic coupling [10].

As introduced in Fig. 5, we weigh the difference images with the corresponding sensitivity maps (Fig. 4 (e) and (f)) and determine $\Delta R_{\text{IM}}$ as an estimate of the expected relative change in the transverse SMR signal. The magnetic changes associated to the pulses in Fig. 6 (a), (b) and (c) result in a stronger increase (a), slight increase (b) and reduction (c) of $\Delta R_{\text{IM}}$, as visible in the weighted difference images in the last column of Figs. 6 (a) – (c).

Turning towards the actual measurements of the transverse resistance, we can observe similar trends in the step height $\Delta R_{\text{T}}$, which is displayed in Fig. 6 (d) after subtracting a linear non-magnetic contribution for each pulse direction. This is a good indication for a magnetic origin of the thus analyzed electrical data.

To further substantiate the comparison of the imaging and electrical data, we expanded the analysis to measurements with various pulse current densities in each of the pulse configurations introduced in Figs. 6 (a) – (c). For each individual pulse, Fig. 6 (e) includes the processed electrical data divided by the average longitudinal resistance $\overline{R}_{\text{L}}$ of the corresponding device, as a function of $\Delta R_{\text{IM}}$. A good correlation over multiple pulse current densities holds not only in the individual measurement geometries, but also with approximately equal slopes in all configurations. These consistent results across the different device geometries, especially with respect to the bidirectional switching patterns observed in the x-shaped pulse geometry (two-level contrast in the difference image of Fig. 6 (c)) and the off-center switching in Fig. 6 (b), strongly support a magnetic origin of the electrical data, processed according to the subtraction procedure introduced in reference [9]. The deviation of the red square data (straight pulses) from a linear slope in Fig. 6 (e) reflects the limitations of the subtraction procedure. At elevated pulse current densities, non-magnetic signals exhibit increased non-linearity, hence the subtraction of a linear function cannot fully account for the non-magnetic contributions and a residual signal remains in the processed data.

**Conclusions**

To conclude, we developed a method to simulate the magnetic sensitivity distribution of structures designed to measure transverse voltage signals arising from the Hall effect or related magnetoresistance effects. In contrast to a symmetric sensitivity in established *Hall cross* structures [22], eight-terminal *Hall star* devices with readout channels of reduced width show a highly localized and asymmetric sensitivity distribution. Our simulations indicate that the magnetic sensitivity can a priori neither be assumed symmetric nor maximized in the device center when utilizing recently established or new device geometries [17,31]. In general, it may be beneficial to use highly symmetric devices with channels of equal width to achieve a relatively uniform magnetic sensitivity in the device center. In view of current-induced domain switching experiments of AFMs, where such structures are frequently used, this result indicates that magnetic contributions to the transverse resistance readout vary in case of irregular domain switching. It is thus of utmost importance to achieve highly regular switching of large domains to ensure a significant and reproducible magnetic signal in the electrical transverse resistance readout. For the example of NiO/Pt devices which fulfill these requirements, we utilize the knowledge of the corresponding sensitivity distributions to translate domain switching patterns observed by direct imaging to an expectation of the associated change in the SMR signal. This estimate correlates with the actual transverse resistance measurements after subtraction of a linear contribution from the electrical data, as discussed before [9]. This consistency strongly supports the presence of a magnetic SMR signal in the transverse resistance changes associated with current-induced switching of the antiferromagnetic Néel vector. Our results can thus motivate further research to reduce non-magnetic signals in the electrical readout and enable the reliable, all-electrical control of the antiferromagnetic Néel order, which is key to develop future spintronic devices based on AFMs.

**Acknowledgments**

L.B acknowledges the European Union's Horizon 2020 research and innovation program under the Marie Skłodowska-Curie grant agreements ARTES number 793159. L.B. and M.K. acknowledge support from the Graduate School of Excellence Materials Science in Mainz (MAINZ) DFG 266, the DAAD (Spintronics network, Project No. 57334897 and Insulator Spin-Orbitronics, Project No. 57524834) and all groups from Mainz acknowledge that this work was funded by the Deutsche Forschungsgemeinschaft (DFG, German Research Foundation) - TRR 173 – 268565370 (projects A01, A03, A11, B02, and B12) and KAUST (OSR-2019-CRG8-4048). R.R. also acknowledges support from the European Commission through Project No. 734187-SPICOLOST (H2020-MSCA-RISE-2016), the European Union's Horizon 2020 research and innovation program through Marie Skłodowska-Curie Actions Grant Agreement SPEC No. 894006, the MCIN/AEI (RYC 2019-026915-I), the Xunta de Galicia (ED431B 2021/013, Centro Singular De Investigación de Galicia Accreditation 2019-2022, ED431G 2019/03) and the European Union (European Regional Development Fund - ERDF). M.K. acknowledges financial support from the Horizon 2020 Framework Programme of the European Commission under FET-Open grant agreement no. 863155 (s-Nebula). This work was also supported by ERATO "Spin Quantum Rectification Project" (Grant No. JPMJER1402) and the Grant-in-Aid for Scientific Research on Innovative Area, "Nano Spin Conversion Science" (Grant No. JP26103005), Grant-in-Aid for Scientific Research (S) (Grant No. JP19H05600), Grant-in-Aid for Scientific Research (C) (Grant No. JP20K05297) from JSPS KAKENHI, Japan.

## Author contributions

L.B. and M.K. proposed and supervised the project. F.S. performed the simulations with inputs from H.M. and carried out the experiments. The devices were designed and fabricated by L.B. and H.M. The antiferromagnetic thin film bilayers were grown by C.S. and R.R. with inputs from L.B and supervised by E.S. F.S. wrote the paper with L.B., H.M. and M.K. All authors commented on the manuscript.

## Data availability

The data that support the findings of this study will be openly available in Zenodo (including a DOI) upon acceptance of the manuscript.